\documentclass[aps,prl,reprint,showpacs,groupedaddress]{revtex4-1}

\usepackage{amsmath}
\usepackage{amssymb}
\usepackage{graphicx}
\usepackage[capitalize]{cleveref}
\usepackage{color}

\def\pl{\partial}

\def\al{\alpha}
\def\bt{\beta}
\def\Ga{\Gamma}

\def\de{\delta}
\def\De{\Delta}

\def\si{\sigma}
\def\Si{\Sigma}
\def\te{\theta}

\def\La{\Lambda}
\def\lam{\lambda}

\def\om{\omega}

\def\sq{\sqrt}

\def\l{\left (}
\def\r{\right )}

\def\fr{\frac}

\def\hs{\hspace}
\def\vs{\vspace}

\def\ran{\rangle}
\def\lan{\langle}
\def\ov{\overline}

\def\tm{\times}

\begin{document}

\newcommand{\ba}[1]{\begin{array}{#1}} \newcommand{\ea}{\end{array}}



\def\Journal#1#2#3#4{{#1} {\bf #2}, #3 (#4)}

\def\NCA{\em Nuovo Cimento}
\def\NIM{\em Nucl. Instrum. Methods}
\def\NIMA{{\em Nucl. Instrum. Methods} A}
\def\NPB{{\em Nucl. Phys.} B}
\def\PLB{{\em Phys. Lett.}  B}
\def\PRL{\em Phys. Rev. Lett.}
\def\PRD{{\em Phys. Rev.} D}
\def\ZPC{{\em Z. Phys.} C}

\def\st{\scriptstyle}
\def\sst{\scriptscriptstyle}
\def\mco{\multicolumn}
\def\epp{\epsilon^{\prime}}
\def\vep{\varepsilon}
\def\ra{\rightarrow}
\def\ppg{\pi^+\pi^-\gamma}
\def\vp{{\bf p}}
\def\ko{K^0}
\def\kb{\bar{K^0}}
\def\al{\alpha}
\def\ab{\bar{\alpha}}

\def\np{Nucl. Phys. {\bf B}}\def\pl{Phys. Lett. {\bf B}}
\def\mpl{Mod. Phys. {\bf A}}\def\ijmp{Int. J. Mod. Phys. {\bf A}}
\def\cmp{Comm. Math. Phys.}\def\prd{Phys. Rev. {\bf D}}

\def\oa{\bigcirc\!\!\!\! a}
\def\ob{\bigcirc\!\!\!\! b}
\def\oc{\bigcirc\!\!\!\! c}
\def\oi{\bigcirc\!\!\!\! i}
\def\oj{\bigcirc\!\!\!\! j}
\def\ok{\bigcirc\!\!\!\! k}
\def\ve{\vec e}\def\vk{\vec k}\def\vn{\vec n}\def\vp{\vec p}
\def\vv{\vec v}\def\vx{\vec x}\def\vy{\vec y}\def\vz{\vec z}

\newcommand{\AdS}{\mathrm{AdS}}
\newcommand{\dd}{\mathrm{d}}
\newcommand{\eee}{\mathrm{e}}
\newcommand{\sgn}{\mathop{\mathrm{sgn}}}

\def\a{\alpha}
\def\b{\beta}
\def\g{\gamma}

\newcommand\lsim{\mathrel{\rlap{\lower4pt\hbox{\hskip1pt$\sim$}}
    \raise1pt\hbox{$<$}}}
\newcommand\gsim{\mathrel{\rlap{\lower4pt\hbox{\hskip1pt$\sim$}}
    \raise1pt\hbox{$>$}}}

\newcommand{\beq}{\begin{equation}}
\newcommand{\eeq}{\end{equation}}
\newcommand{\bea}{\begin{eqnarray}}
\newcommand{\eea}{\end{eqnarray}}
\newcommand{\noi}{\noindent}


\title{Light Pseudo-Goldstone Higgs Boson from $SO(10)$ GUT with Realistic Phenomenology}

\author{Zurab Tavartkiladze}
\email[]{zurab.tavartkiladze@gmail.com}

\affiliation{Center for Elementary Particle Physics, ITP, Ilia State University, 0162 Tbilisi, Georgia}


\begin{abstract}
Within the supersymmetric $SO(10)$ grand unified theory (GUT), a new mechanism, giving the  light Higgs doublet as a pseudo-Goldstone mode, is suggested.
Realizing this mechanism, we present an explicit model with fully realistic phenomenology. In particular, desirable symmetry
breaking and natural all-order hierarchy are achieved.  The constructed model allows one to have a realistic fermion pattern, nucleon stability,
and successful gauge coupling unification. The suggested mechanism opens prospects in the field for a novel $SO(10)$ GUT model building and
for further investigations.
\end{abstract}




\maketitle

\section{I. Introduction}
\vs{-0.3cm}
Despite enormous success, the Standard Model (SM), failing to explain several observations, needs to be extended.
Making steps beyond the SM, the grand unification theory (GUT) \cite{Pati:1973rp}  is a highly motivated and elegant idea.
The best candidate for GUT model building
is the $SO(10)$ group \cite{Georgi:1974my}, whose spinorial $16_i$-plets ($i=1,2,3$), together with SM matter, include
 right-handed neutrino  $\nu^c$ states
(one per generation) -- $16_i=\{ q, u^c, d^c, l, e^c, \nu^c \}_i$.
Thus, within the $SO(10)$ GUT, the most compelling seesaw neutrino mass generation mechanism \cite{see-saw}
finds a natural realization, accommodating the data from various neutrino oscillation experiments.
As shown, the same mass generation mechanism can generate  the observed baryon asymmetry
through leptogenesis \cite{Fukugita:1986hr}.
Towards circumventing other problems, the supersymmetric (SUSY) extensions turn out to be a relief.
SUSY ensures the stability of the weak scale. The SM's minimal SUSY extension (MSSM)
  leads to successful gauge coupling unification \cite{Dimopoulos:1981yj} (a necessary ingredient for GUTs)
 and  includes the viable dark matter candidate -- the neutralino.

Although there are numerous merits of the SUSY $SO(10)$ GUT (we study henceforth), the most challenging problem --
the doublet-triplet (DT) splitting problem -- (together with other puzzles we address below) needs to be taken seriously \cite{DT-problem}.
The missing vacuum expectation value (VEV) mechanism, proposed earlier \cite{Dimopoulos:1981xm}, enabling one to have quite realistic constructions  \cite{Babu:1993we,Berezhiani:1996bv,Barr:1997hq,Babu:1998wi,Babu:2010ej,raby}, triggered an intense investigation in the
field \cite{raby}.
An alternative, and perhaps most elegant, idea for resolving the DT problem is the pseudo-Goldstone boson (PGB) mechanism, which proposes
 the light Higgs as the PGB \cite{Inoue:1985cw,Berezhiani:1989bd}.
 However, so far, attempts to realize this idea within GUTs,  have been successful only within $SU(6)$ gauge symmetry
 \cite{Berezhiani:1989bd,pgb-su6,Dvali:1996sr}.
 Obviously, it would be very exciting to find a possibility for the PGB mechanism realization within the $SO(10)$ GUT.

 In this paper, the first time within the SUSY $SO(10)$ GUT, I propose the possibility for the Higgs doublet supermultiplets emerging as light
 PGBs. The model, I present, turns out to be fully realistic. Consistent GUT symmetry breaking is achieved, and the MSSM Higgs doublets are
 light  pseudo-Goldstones,  even after taking into account all allowed high-order operators.
 The model I construct  offers a realistic fermion pattern, nucleon stability,
and successful gauge coupling unification. Because of these issues (usually turning out to be severe problems for various GUTs), I find the presented $SO(10)$ model to be quite successful.

The mechanism proposed in this paper opens the door for future work to build varieties of PGB $SO(10)$ models and investigate their phenomenological implications.

\section{II. The Setup and the Mechanism}
\vs{-0.3cm}
First, I describe the PGB mechanism within the SUSY $SO(10)$ GUT and then, on a concrete model, demonstrate its
natural realization.
The minimal superfield content, ensuring $SO(10)\to SU(3)_c\tm SU(2)_L\tm U(1)_Y\equiv G_{\rm SM}$ breaking, is $\{ 45_H$, $16_H$, $\ov{16}_H\} $ [and possibly some additional $SO(10)$ singlet states]. Thus, the symmetry-breaking scalar superpotential  $W_H$ consists of three parts:
\begin{eqnarray}
\label{WH}
&W_H=W_H^{(45)}+W_H^{(16)}+W_H^{(45,16)}~,~~~~
\end{eqnarray}
where $W_H^{(45)}$ and $W_H^{(16)}$ involve separately $45_H$ and $\{16_H, \ov{16}_H\}$-plets, respectively.
The $W_H^{(45)}$, including some powers of $45_H$ (such as $45_H^2, 45_H^4,\dots $),
is responsible for fixing the VEV $\lan 45_H\ran $ towards the desirable direction.
The $W_H^{(16)}$  fixes the VEVs of $16_H$ and $\ov{16}_H$ in their $\nu^c_H, \ov{\nu }^c_H$ scalar components and
can be $W_H^{(16)}\!=\!S(i\ov{16}_H16_H+v_R^2 )$, where $S$ is the $SO(10)$ singlet. The condition $F_S=0$,
together with $D$-flatness, would fix $\lan \nu^c_H\ran =\lan \ov{\nu }^c_H\ran =v_R $. For us, it is essential that the terms like $(\ov{16}_H16_H)^2$ and $\ov{16}_H16_H45_H^{2n+1}$
are absent or are properly suppressed (will be explained below).  The superpotential $W_H^{(45,16)}$ includes couplings of $\ov{16}_H16_H$
with {\it only even} powers of $45_H$ (this can be guaranteed by discrete symmetries), such as $\ov{16}_H45_H^216_H$.
Without $W_H^{(45,16)}$, the $W_H$ has global symmetry $SO(10)_{45}\tm U(16)$ [e.g., $W_H^{(45)}$ possesses $SO(10)_{45}$ symmetry, while
the $W_H^{(16)}$  has $U(16)$ global symmetry]. With the VEVs $\lan \nu^c_H\ran $ and $\lan \ov{\nu }^c_H\ran $,
the breaking $U(16)\to U(15)$ happens and pseudo-Goldstone states emerge.
Clearly, those modes (at least some of those) would gain masses by including the terms of $W_H^{(45,16)}$. Our goal is to see if it is possible
to arrange the couplings such that MSSM Higgs doublets emerge as light pseudo-Goldstones while all remaining  states acquire large masses.
It turns out that in this setup this is indeed possible.

Assume that the $45_H$ has a VEV towards the $B-L$ direction $\lan 45_H\ran \sim V_{\!B\!-\!L}$.
 Thus, with $SO(10)\to SU(4)_c\tm SU(2)_L\tm SU(2)_R$ decomposition
\begin{eqnarray}
\label{45-422-dec}
&45_H=\Si (15,1,1)+\Si_R(1,1,3)+\Si_L(1,3,1)+(6,2,2) ,~~~~
\end{eqnarray}
the $SU(4)_c$'s adjoint $\Si (15,1,1)$ have the VEV
$\lan \Si \ran =V_{\!B\!-\!L}\!\cdot \!{\rm Diag}\l 1, 1, 1, -3\r $,
causing the breaking of $SO(10)$ gauge symmetry down to $G_{3221}\equiv SU(3)_c\tm SU(2)_L\tm SU(2)_R\tm U(1)_{\!B\!-\!L}$.
With this breaking, the $30$ states (from $45_H$) absorbed by superheavy gauge fields,  become genuine Goldstones. These are the fragment $(6,2,2)$ [given in (\ref{45-422-dec})]
plus $SU(3)_c$'s triplet-antitriplet pair $3_{\Si }, \bar 3_{\Si }$ coming from $\Si (15,1,1)$. The remaining fragments of $45_H$ should gain large masses. Moreover, with $W_H^{(16)}$'s terms, such as $\lam_S S(i\ov{16}_H16_H+v_R^2)$,
one can have $\lan \nu^C_H\ran = \lan \ov{\nu }^C_H\ran =v_R$, inducing the breaking
 $G_{3221}\to G_{\rm SM}$.  The spontaneous breaking $W_H^{(16)}$'s  global $U(16)$
would give unwanted pseudo-Goldstone modes. However, the superpotential $W_H^{(45,16)}\supset \fr{\lam i}{4!M_{*}}\ov{16}_H45_H^216_H$
coupling will break $U(16)$ but, due to the $\lan 45_H\ran $ configuration, render some global symmetries.
In particular, using in this coupling the VEV $\lan 45_H\ran =\lan \Si \ran \propto Q_{\!B-L\!}$
[see Eq. (\ref{45-422-dec}) and the comment below it],
the bilinear terms with respect to the states $16_H=\{ f_H\}, \ov{16}_H=\{ \bar f_H\}$, where
$f_H=\{ {\nu}^c_H, e^c_H, l_H, u^c_H, d^c_H, q_H\}$,
will be
\begin{eqnarray}
\label{U4-U12}
&\fr{\lam }{M_{*}}\ov{16}_H45_H^216_H
\to \lam \fr{V_{\!B\!-\!L}^2}{M_*}\sum_f\bar f_H(Q^f_{B-L})^2f_H+\cdots
 \nonumber \\
&=\lam \fr{V_{\!B\!-\!L}^2}{M_*}\l \bar{\nu}^c_H{\nu}^c_H+\bar e^c_He^c_H+\bar l_Hl_H\r   \nonumber  \\
&~~~+\lam \fr{V_{\!B\!-\!L}^2}{M_*}\fr{1}{9}\l \bar u^c_Hu^c_H+\bar d^c_Hd^c_H+\bar q_Hq_H\r +\cdots ,
\end{eqnarray}
where I have displayed terms  not respecting the $U(16)$ global symmetry.
However, due to the relation
 $M_{f_H\bar f_H}\propto (Q_{\!B-L \!}^f)^2$, as  seen from (\ref{U4-U12}), the second and last lines possess $U(4)$ and  $U(12)$ symmetries, respectively. The $U(4)$ global symmetry, which has leptonic states, is spontaneously broken down to $U(3)$ by the VEVs
 $\lan \nu^C_H\ran = \lan \ov{\nu }^C_H\ran =v_R$.
  Because of $U(4)\to U(3)$ breaking the seven Goldsones emerge. Among them, three -- one SM singlet and the pair $e^c_H, \bar e^c_H$ -- are
 genuine Goldstones absorbed by the coset $[SU(2)_R\tm U(1)_{\!B-L\!}]/U(1)_Y$ gauge fields.
 The states $l_H$ and $\bar l_H$, having the
 quantum numbers of the MSSM Higgs superfields, emerge as light (as desired) pseudo-Goldstone modes.  The remaining fragments,
 given in the last line in (\ref{U4-U12}), will be heavy.

However, for this mechanism to be realized, some care needs to be exercised. The $\lam $ coupling term, with nonzero
$\lan \nu^C_H\ran = \lan \ov{\nu }^C_H\ran =v_R$, would also trigger the VEV of $45_H$ in the $I_{3R}$ direction, i.e., the
VEV of $\Si_R(1,1,3)$ of (\ref{45-422-dec}) with the
$\lan \Si_R \ran =V_{R}\!\cdot \!{\rm Diag}\l 1, -1\r $
configuration in $SU(2)_R$ space. As expected, $V_R\sim \lam v_R^2/V_{\!B\!-\!L}$. This, in (\ref{U4-U12}), would introduce $U(4)$ symmetry-breaking
mass terms$\sim \lam^2v_R^2/M_*$. Thus, the PGB's mass -- the $\mu $-term -- will be $\mu \sim \lam^2v_R^2/M_*$.
For its proper suppression, for $v_R\!=\!10^{16-17}$~GeV, $M_*\!\simeq \!2.4\cdot 10^{18}$~GeV, we need
$\lam \stackrel{<}{_\sim } 10^{-6}$.
Besides this, terms such as $\ov{16}_H45_H^{2n+1}16_H$  and $(16_H\ov{16}_H)^{k+2}$ ($n, k=0,1,2, \dots $) should be absent or adequately suppressed.
Below, I present the model where all this is naturally realized; i.e., smallness or absence of any coupling will follow from the
symmetries (we will invoke).

\section{III. Model with All-Order Natural Hierarchy}
\vs{-0.3cm}
Realizing the mechanism described above, we augment the SUSY $SO(10)$
GUT by  ${\cal U}(1)_A\tm {\cal Z}_4$ symmetry, where ${\cal U}(1)_A$ is an anomalous gauge symmetry and ${\cal Z}_4$ is
discrete $R$ symmetry. For symmetry breaking and all-order DT hierarchy, the anomalous ${\cal U}(1)_A$
was first applied in Ref. \cite{Dvali:1996sr}  [within the PGB $SU(6)$ GUT] and was proven to be  very efficient for realistic $SO(10)$ model building \cite{Berezhiani:1996bv,Babu:2010ej}.
Besides the states $45_H, 16_H$ and  $\ov{16}_H$,
we introduce two $SO(10)$ singlet superfields  $S$ and  $S_1$, which will be used in superpotential $W_H$ of Eq. (\ref{WH}).
In Table \ref{t:U-Z-charges}, I display the ${\cal U}(1)_A\tm {\cal Z}_4$ transformation properties of these fields
together with other states (introduced and discussed later on).
Note that, under ${\cal Z}_4$ symmetry, the whole superpotential transforms as $W\to -W$.
%
%
%
%
\begin{table}
  \caption{\label{t:U-Z-charges}
${\cal U}(1)_A$ and ${\cal Z}_4$ charges $Q_i$ and $\om_i$ of the superpotential and superfield $\phi_i$. The transformations under
${\cal U}(1)_A$ and ${\cal Z}_4$ are, respectively, $\phi_i\to e^{iQ_i}\phi_i$ and $\phi_i\to e^{i\om_i}\phi_i$; $\om =\fr{2\pi }{4}$.
}
\begin{center}
\begin{tabular}{|l|c|c|c|c|c|c|c|c|c|}
\hline
{\rule{0mm}{4mm}}%
 &$W$  & $45_H$ & $16_H$  & $\ov{16}_H$ & ~$S$~ & ~$S_1$ ~& $10_H$ & ${10'}_H$ & $16_i$   \\
\hline
{\rule{0mm}{5mm}} $Q_i$ & $0$ & $0$ & $-5$ & $1$ & $0$ & $4$ & $-6$ & $6$ & $3$  \\
\hline
{\rule{0mm}{5mm}} $\om_i$ & $2\om $ & $\om $ & $-\om $ & $\om $ & $2\om $ & $0$ & $0$ & $\om $ & $\om $  \\
\hline
\end{tabular}
\end{center}
\end{table}
%
%
%
%
%
Thus, the relevant superpotential couplings  are
\begin{eqnarray}
\label{WH-model}
&W_H^{(45)}=\fr{1}{2}M_{45}{\rm tr}(45_H^2)+\fr{\lam_1}{M_{*}^3}[{\rm tr}(45_H^2)]^3
 \nonumber \\
&+\fr{\lam_2}{M_{*}^3}{\rm tr}(45_H^2){\rm tr}(45_H^4) +\fr{\lam_3}{M_{*}^3}{\rm tr}(45_H^6) \nonumber \\
&W_H^{(16)}=-\lam_S S(i\fr{S_1}{M_*}\ov{16}_H16_H+\La^2),  \nonumber \\
&W_H^{(45,16)}=\fr{\bar \lam iS_1}{4!M_{*}^2}\ov{16}_H45_H^216_H .
\end{eqnarray}
With anomalous ${\cal U}(1)_A$, having the string origin,  the Fayet-Iliopoulos term  $\int d^4\te \xi V_A$ is always
generated \cite{Dine:1987xk}, and the corresponding $D$-term potential is
\begin{eqnarray}
\label{VD}
&V_D\!=\!\fr{g_A^2}{8}(\xi \!+\!\sum_iQ_i|\phi_i|^2)^2\!\to \!\fr{g_A^2}{8}(\xi -4v_R^2+4|\lan S_1\ran |^2)^2 .~~~~
\end{eqnarray}
At the last stage in (\ref{VD}), we have taken into account that  $\lan \nu^c_H\ran =\lan \bar{\nu }^c_H\ran \equiv v_R$,
which ensures the vanishing of all $D$ terms of $SO(10)$.  With $\xi >0$ and looking for the solution of $\lan S_1\ran \ll v_R$,
the condition $\lan D_A\ran =0$ from (\ref{VD}) gives
\begin{eqnarray}
\label{fix-vR}
&v_R\simeq 0.5\sq{\xi }~ .~~~
\end{eqnarray}
Next, we investigate the symmetry-breaking pattern from the superpotential couplings given in (\ref{WH-model}).
With the branching $\ov{16}\tm 16\!=\!1+45+210$ and noting that $45_H^2$ identically vanishes in the antisymmetric $45$ channel, the
remaining relevant contraction [breaking global $U(16)$ symmetry] is in the $210$ channel. Thus, under the $\ov{16}_H45_H^216_H$ term, we assume the invariant
$\l \ov{16}_H16_H\r_{\!210}\!\!\cdot \!\l 45_H^2\r_{\!210}$ (see the Appendix for more details).

Writing $45_H$'s VEV in $SO(10)$ group space as
\begin{eqnarray}
\label{45-VEV}
&\lan 45_H\ran =i\si_2\otimes {\rm Diag}\l V_{\!B\!-\!L}, V_{\!B\!-\!L}, V_{\!B\!-\!L}, V_{\!R}, V_{\!R}\r ~~~
\end{eqnarray}
and imposing $F_{\phi_i}=0$,
from the superpotential couplings (\ref{WH-model}), taking into account that $v_R$ is fixed as (\ref{fix-vR}),  we obtain
\begin{eqnarray}
\label{sol-S1-S}
&\lan S_1\ran \simeq \fr{\La^2}{v_R^2}M_* ,~~
\lan S\ran =-\fr{3\bar \lam }{\lam_SM^*}V_{\!B\!-\!L}^2(1+\fr{2V_{\!R}}{V_{\!B\!-\!L}}+\fr{V_{\!R}^2}{3V_{\!B\!-\!L}^2}),~~~~~
\end{eqnarray}
\begin{eqnarray}
\label{eq-VBL}
&{\small 6(36\lam_1\!+\!6\lam_2\!\!+\!\!\lam_3)\!V_{\!B\!-\!L}^4\!\!+\!\!16(18\lam_1\!\!+\!\!\lam_2)\!V_{\!B\!-\!L}^2\!V_{\!R}^2\!\!+
\!\!8(12\lam_1\!\!+\!\!\lam_2)\!V_{\!R}^4}
 \nonumber \\
&=-M_{45}M_*^3+\bar \lam \lan S_1\ran v_R^2M_*\l 1+\fr{V_{\!R}}{V_{\!B\!-\!L}}\r ,
\end{eqnarray}
\begin{eqnarray}
\label{eq-VR}
&{\small (16\lam_1\!+\!4\lam_2\!\!+\!\!\lam_3)\!V_{\!R}^5\!\!+\!\!4(12\lam_1\!\!+\!\!\lam_2)\!V_{\!B\!-\!L}^2\!V_{\!R}^3\!\!+
\!\!2(18\lam_1\!\!+\!\!\lam_2)\!V_{\!B\!-\!L}^4\!V_{\!R}}
 \nonumber \\
&+\fr{1}{6}M_{45}M_*^3V_{\!R}=\fr{1}{4}\bar \lam \lan S_1\ran v_R^2\!V_{\!B\!-\!L}M_*\l 1+\fr{V_{\!R}}{3V_{\!B\!-\!L}}\r .~~~~
\end{eqnarray}
From (\ref{sol-S1-S})--(\ref{eq-VR}), one can see that the desirable VEV configuration is obtained.
Indeed, from (\ref{sol-S1-S}), with $v_R\sim 10^{17}$~GeV [fixed from (\ref{fix-vR})] and $\La =(10^{13}-10^{14})$~GeV, we obtain $\fr{\lan S_1\ran }{M_*}\simeq 10^{-8}\!\!-\!\!10^{-6}$. Thus, an effective, and suppressed, $\lambda $ coupling (ensuring suppressed Higgs mass) is generated.
Small $\lan S_1\ran $ also ensures naturally suppressed $V_R$
(in a limit $\lan S_1\ran \to 0$, one has the solution $V_R\to 0$). At the leading order, in powers of $\fr{\lan S_1\ran }{M_*}$,
from (\ref{sol-S1-S})--(\ref{eq-VR}) we get
\begin{eqnarray}
\label{approx-sols}
&{\rm with}~~\fr{\La }{v_R}\ll 1:~
 V_{\!B\!-\!L}\!\simeq \!\!\l \!\fr{-M_{45}M_*^3}{6(36\lam_1\!+\!6\lam_2\!+\!\lam_3)}\!\r^{1/4}\!\!\!\!,~~
 \fr{\lan S_1\ran }{M_*}\simeq \fr{\La^2}{v_R^2} ,
 \nonumber \\
& V_{\!R}\simeq -\bar \lam \fr{\lan S_1\ran}{M_*} \fr{v_R^2M_*^2}{4(4\lam_2+\lam_3)V_{\!B\!-\!L}^3},~~~
 \lan S\ran \simeq -\fr{3\bar \lam }{\lam_S}\fr{V_{\!B\!-\!L}^2}{M_*} .
\end{eqnarray}

For the masses of $l_H$ and $\bar l_H$, coming from the $16_H$ and  $\ov{16}_H$, we have
(in the Appendix, I present the decomposition helping to compute the masses of this and remaining fragments)
\begin{eqnarray}
\label{M-lH}
&M_{\bar l_H\!l_H}\!\!=\!\fr{-6\bar \lam \lan S_1\ran V_{\!B\!-\!L}V_{\!R}}{M_*^2}\!( \!1\!-\!\fr{V_{\!R}}{3 V_{\!B\!-\!L}} \! ) \!
\simeq \!\!\l \!\fr{\La }{v_R}\!\r^{\!\!4}\!\!\!\fr{3\bar \lam^2v_R^2M_*}{2(4\lam_2\!+\!\lam_3)V_{\!B\!-\!L}^2}. ~~~~
\end{eqnarray}
The latter's value, i.e., the $\mu $ term, with $v_R\sim V_{\!B\!-\!L}$ and  for
$\fr{\La }{v_R}\!\!\sim \!\!(1\!-\!2)\!\!\cdot \!\!10^{-4}$
will be  $\!\sim \!1\!-\!10$~TeV -- of desirable magnitude.
 Allowed higher-order operator $\fr{1}{M_*^5}(S_1\ov{16}_H45_H16_H)^2$ is harmless, because  it contributes to the
 $\mu $ term by the amount $\sim \fr{\lan S_1\ran^2V_{\!B\!-\!L}^2v_R^2}{M_*^5}\stackrel{<}{_\sim } 10^{-2}$~GeV
(for $V_{\!B\!-\!L} , v_R \sim M_*/30$).

The  masses of colored modes (from $16_H, \ov{16}_H$) are
\begin{eqnarray}
\label{M-T1}
&M_{\bar q_H\!q_H}\!\simeq \!M_{\bar u^c_H\!u^c_H}\!\simeq
\!M_{\bar d^c_H\!d^c_H}\!\!\equiv \!M_{T_1}\!\simeq  \!-4\bar \lam \!\l \!\fr{\La }{v_R}\!\r^{\!\!2}\!\fr{V_{\!B\!-\!L}^2}{M_*} ~,~~
\end{eqnarray}
around $\sim 10^8$~GeV.

Below, I show how MSSM Higgs doublets can have desirable couplings to the MSSM matter and how
nucleon stability is ensured \cite{tripl-mass}.
Before discussing these, I give  the scalar sector's extension, which
ensures desirable Yukawa couplings and realistic phenomenology. I introduce two scalar superfields in the fundamental representation ($10$-plets) of $SO(10)$: $10_H$ and $10_H^{\hs{-0.1mm}'}$ with
  ${\cal U}(1)_A\tm {\cal Z}_4$ charges given in Table \ref{t:U-Z-charges}.
Thus, the relevant superpotential couplings are
\begin{eqnarray}
\label{W-10-16}
&W_H^{(16,10)}=\al_1\fr{S_1}{M_*}\ov{16}_H\ov{16}_H10_H+\lam'10_H45_H{10'}_H ~~~
 \nonumber \\
& +\al_2\fr{(\ov{16}_H16_H)^3}{M_*^5}{10'}_H{10'}_H ~~~~~
\end{eqnarray}
($\al_{1,2}, \lam'$ are dimensionless couplings), which, with notations $\{l_H, \bar l_H\}\!\equiv \!\{h_d^{16_H}\!, h_u^{\ov{16}_H}\}$,
give the doublet mass matrix:
\begin{eqnarray}
&
\begin{array}{ccc}
 & {\begin{array}{ccc}
\hs{-1.8cm} h_d^{16_H}&  h_d^{10_H}&  h_d^{{10'}_{\!\!H}}
\end{array}}\\ \vspace{1mm}
M_D= \hs{-0.2cm}
\begin{array}{c}
h_u^{\ov{16}_H} \\ ~ h_u^{10_H} \\~h_u^{{10'}_{\!\!H}}
 \end{array}\!\!\!\!\!\hs{- 0.1cm}&{\left(\begin{array}{ccc}

 0& \bar M & 0
\\
0&0 &\lam'V_R
 \\
 0& -\lam'V_R & M'
\end{array}\right)}+{\cal O}({\rm TeV})~,
\end{array}  \!\! \label{M-doubl} \\
&{\rm with}~~~~~\bar M=\al_1\fr{\lan S_1\ran v_R}{M_*} ~,~~~M'=\al_2\fr{v_R^6}{M_*^5}~.\label{M-entries}
\end{eqnarray}
 In (\ref{M-doubl}), the ${\cal O}({\rm TeV})$ stand for possible corrections of the order of $(1\!\!-\!\!10)$TeV (I will comment shortly) and
are harmless.

 From (\ref{M-doubl}), we see that one
  doublet pair, identified with MSSM Higgs superfields $h_u$ and  $h_d$, is light. The remaining doublets
   (denoted as $D_{1,2}$ and $\bar D_{1,2}$) are heavy.
 Analyzing the matrix (\ref{M-doubl}), the distribution of  $h_u$ and $h_d$ in original states can be found:
 \begin{eqnarray}
\label{h-weights}
&h_u^{10_H}=e^{i\phi_x}|x|h_u+\cdots ,~~~~h_d^{16_H}=h_d ,
 \nonumber \\
&{\rm with}~~\fr{1}{|x|}=\fr{1}{|M'|}\!\l |M'|^2\!+\!|\lam'V_R|^2\!+\!\fr{|\lam'V_R|^4}{|\bar M|^2}\r^{1/2}~ ~~~
\end{eqnarray}
the composition crucial for building a realistic fermion pattern (especially for large top Yukawa coupling $\lam_t$).

As far as other operators, allowed by symmetries, are concerned,
the couplings
$\fr{SS_145_H}{M_*^3}16_H16_H{10'}_{\!\!H}$ and $\fr{S45_H}{M_*^6}(\ov{16}_H16_H)^2\ov{16}_H\ov{16}_H{10'}_{\!\!H}$
with $\fr{\lan S\ran }{M_*}\!\sim \!10^{-4}, \fr{\lan S_1\ran }{M_*}\!=\!\fr{\La^2}{v_R^2}\!\sim \!10^{-8}$,
$v_R , V_{\!B\!-\!L}\!\sim \!\fr{M_*}{30}$,
will contribute to the $\mu $ term by an amount of $\sim {\rm few~ TeV}$ and therefore are harmless \cite{harmless}.

\section{IV. Yukawa Couplings and Proton Stability}
\vs{-0.3cm}
All MSSM matter is embedded in  $SO(10)$'s $16_i$-plets ($i=1,2,3$). Their ${\cal U}(1)_A\!\times \!{\cal Z}_4$
charges are given in Table \ref{t:U-Z-charges}. The effective operators, generating up- and down-type quark and charged lepton masses, are
 \begin{eqnarray}
\label{yuk-ops}
&W_Y\!=\!Y_U^{ij}\!16_i16_j10_H+\fr{45_H^2}{\lan \ov{16}_H16_H\ran }\!\!\fr{Y^{ij}}{M_*}16_i16_j16_H16_H .~~~~~~
\end{eqnarray}
The first term is responsible for the up-type quark Yukawa couplings.
 According to (\ref{h-weights}), the $10_H$-plet includes the $h_u$ by the weight $|x|$, which
  with $|\lam' V_R|\sim |\bar M|\sim |M'|$ can naturally be $|x|\approx 0.5-1$.
  Therefore, the $\lam_t$ coupling  can naturally have a desirable value.
 On the other hand, the $h_d$ entirely resides in the $16_H$-plet, and down quark and charged lepton masses
 are emerging from the second term of (\ref{yuk-ops}),
 which can be obtained by integrating out heavy vectorlike fermion superfields. For instance, with additional  $10_f$ states
 [with ${\cal U}(1)_A\tm {\cal Z}_4$ charges $Q_{10_f}=2, \om_{10_f}=\om $]
 and couplings
 $[\fr{y^{i\al}}{M_*}16_H45_H16_i10_{f\al } -\fr{\kappa^{\al \bt }}{4M_*}\ov{16}_H16_H10_{f\al }10_{f\bt }]$,
one can  see that integration of $10_f$ states (gaining masses after substitution of the VEV $\lan \ov{16}_H16_H\ran $) leads to the second term of (\ref{yuk-ops}) with
$Y^{ij}\simeq (y\kappa^{-1}y^T)^{ij}$. Note that, the $45_H^2$  can couple to the third family by the
contraction ${\rm tr}(45_H^2)$ [i.e., $\fr{{\rm tr}(45_H^2)}{\lan \ov{16}_H16_H\ran}16_316_316_H16_H $] and, therefore, one can have $\lam_b=\lam_{\tau }$ -- the bottom-tau unification -- at the GUT scale.
The couplings  $y^{i\al}$ and $\kappa^{\al \bt }$ can also be obtained via renormalizable interactions. For example,
having additional $16_f$ and $\ov{16}_f$ states
[with $Q_{16_f}\!\!=\!\!-Q_{\ov{16}_f}\!=\!3$ and $\om_{16_f}\!=\!2\om, \om_{\ov{16}_f}\!=\!0$]
and superpotential terms $16_i45_H\ov{16}_f+16_H10_f16_f+M_f\ov{16}_f16_f$, decoupling of $16_f$ and $\ov{16}_f$  generates a term with $y^{i\al}$ coupling above.  I will not discuss here more details (should be pursued elsewhere), but emphasize that
 after the GUT symmetry breaking
some fragments from $10_f, 16_f$ and $\ov{16}_f$ can have masses below the GUT scale and their contribution,
as thresholds, may play an important role for the gauge coupling unification (discussed at the end).

The right-handed neutrinos $\nu^c_i$ are gaining masses($\equiv M_R$) via the effective operator
$(16_i\ov{16}_H)^2\fr{(\ov{16}_H16_H)^2}{\lan S\ran M_*^4}$, which can be generated
by decoupling the $SO(10)$ singlets with masses$\sim \lan S\ran $. With  $\fr{\lan S\ran }{M_*}\!\sim \!10^{-4}$,
$M_R\!\sim \!\fr{v_R^6}{\lan S\ran M_*^4}\!\approx \!10^{14}$~GeV  indeed of desirable values, generating the light neutrino masses
$\sim 0.1$~eV via the seesaw mechanism \cite{see-saw}.
Before discussing  $d=5$ proton decay,  note that ${10'}_H$'s couplings with matter
$\fr{1}{M_*^8}(\ov{16}_H16_H)^3S45_H16_i16_j{10'}_{\!\!H}\to \sim 10^{-14}16_i16_j{10'}_{\!\!H}$
are so suppressed that they should be completely ignored.

The couplings in (\ref{yuk-ops}), after substituting appropriate VEVs,  besides the Yukawa interactions
 \begin{eqnarray}
\label{Yuk-eff}
&W_Y\to (qY_Uu^c +l Y_U\nu^c)h_u+(qY_Dd^c+e^cY_El)h_d~,~~~~~~
\end{eqnarray}
also generate couplings
 \begin{eqnarray}
\label{matt-tripl}
&\fr{1}{2}qY_UqT_{10_H}+qY_Ul\bar T_{10_H}+qY_{ql}l\bar T_{16_H} ~~~
\end{eqnarray}
(with notation $\bar T_{16_H}=d^c_H$). Integration of the triplet states, forming the
mass matrix coupling $T^TM_T\ov T$, leads to a baryon number-violating $LLLL$ $d=5$
operator:
 \begin{eqnarray}
\label{d5-LLLL}
&{\cal O}^{LLLL}_{d=5}=\fr{1}{2} [qY_{ql}l(M_T^{-1})_{12}+qY_Ul(M_T^{-1})_{22}] qY_Uq~ .~~~~~
\end{eqnarray}
The triplet mass matrix, derived from the couplings (\ref{W-10-16}) and taking into account  (\ref{M-T1}), has the form
\beq
\begin{array}{ccc}
 & {\begin{array}{ccc}
\hs{-0.3cm} \ov T_{16_H}& \ov T_{10_H}& \ov T_{{10'}_{\!\!H}}
\end{array}}\\ \vspace{1mm}
M_T= \hs{-0.2cm}
\begin{array}{c}
T_{\ov{16}_H} \\ ~T_{10_H}  \\~T_{{10'}_{\!\!H}}
 \end{array}\!\!\!\!\!\hs{-0.1cm}&{\left(\begin{array}{ccc}
 M_{T_1} & \bar M & 0 \\
0&0 &\lam'V_{\!B\!-\!L}
 \\
 0& -\lam'V_{\!B\!-\!L} & M'
\end{array}\right)}~,
\end{array}  \!\!
\label{M-Tripl}
\eeq
from which
$(M_T^{-1})_{12}\!=\!\fr{-\bar MM'}{M_{T_1}(\lam'V_{\!B\!-\!L})^2}$ and
$(M_T^{-1})_{22}\!=\! \fr{M'}{(\lam'V_{\!B\!-\!L})^2}$.
%
%
For  $\lam'\simeq 1$,  the mass scales $M_{T_1}\simeq 10^8$~GeV, $\bar M\simeq 3\cdot 10^{11}$~GeV and $M'\simeq 8\cdot \!10^{11}$~GeV,  whose values are dictated from (\ref{M-T1}), (\ref{M-entries}) and desirable phenomenology [remember, the
value  $V_R\approx 5\cdot 10^{11}$~GeV obtained from (\ref{approx-sols}) and the need for $|x|\approx 0.5-1$ of (\ref{h-weights}) giving the needed value of $\lam_t$,
so, we take $|\lam'V_R|=5\cdot 10^{11}$~GeV],
 and with $V_{\!B\!-\!L}\simeq 8\!\cdot \!10^{16}$~GeV,  we obtain
$(M_T^{-1})_{12}\!\simeq \!\!\fr{1}{2.7\cdot 10^{18}{\rm GeV}}$ and
$(M_T^{-1})_{22}\!\simeq \!\!\fr{1}{8\cdot 10^{21}{\rm GeV}} $.
These ensure the adequate suppression of $d=5$ operators in (\ref{d5-LLLL}) \cite{sim-suppr}.
Remarkably,  although  the color triplets from $16_H$ and $\ov{16}_H$ have intermediate mass [$\sim 10^8$~GeV; see
Eq. (\ref{M-T1})], the model allows one to have a stabile nucleon and therefore be fully self-consistent.

\section{V. Gauge Coupling Unification.}
\vs{-0.3cm}
\begin{figure}[!t]
\begin{center}
\includegraphics[width=1\columnwidth]{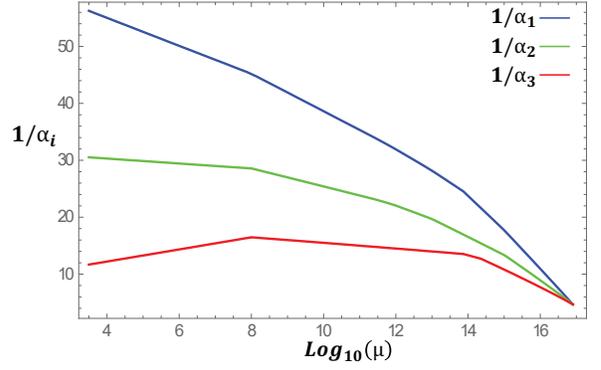}
\caption{Gauge coupling unification within the presented $SO(10)$ model.
The spectrum of states is given in  Eq. (\ref{mas-unif}).
}
\label{SO10-unif-plot}
\end{center}
\end{figure}
Finally, let us discuss the impact of some intermediate states, populating the 'desert' between  the electroweak  and GUT scales.
From (\ref{M-Tripl}), one triplet pair has mass  $M_{T_1}$ given in
(\ref{M-T1}) (together with the remaining colored states).  We denote the masses of the remaining two triplet-antitriplet pairs
by $M_{T_{2,3}}$.
These and heavy doublets' masses $M_{D_{1,2}}$ are found from (\ref{M-Tripl})
and (\ref{M-doubl}): $M_{T_{2,3}}\simeq |\lam' V_{\!B\!-\!L}|$ and
$2M_{D_{1,2}}^2\!=\!|\bar M|^2\!+\!|M'|^2\!+\!2|\lam'V_R|^2\pm \left [ (|\bar M|^2\!-\!|M'|^2)^2+4|\lam'V_RM'|^2\right ]^{1/2}$.
%
%
%
Moreover, from $45_H$ come a colored octet (with mass $M_8$) and $SU(2)_L$ triplet (with mass $M_3$).
One physical $e^c+\bar e^c $ pair, with mass $M_{e^c_H}$, comes
from the superposition of the corresponding fragments from $45_H$ and $16_H, \ov{16}_H$ (the orthogonal combination  is and
absorbed Goldstone). A similar thing happens with the scalar supermultiplets having the quantum numbers of $u^c_H, q_H, \bar u^c_H$ and $\bar q_H$.
We will denote the masses of the corresponding physical states by $M_{u^c_S}$ and $M_{q_S}$.
From Eqs.  (\ref{WH-model}), (\ref{approx-sols}), and (\ref{M-T1}), we have
$\fr{M_8}{8\lam_2+4\lam_3}\simeq \fr{M_{3(e^c_H)}}{4\lam_2+\lam_3}\simeq 6\fr{V_{\!B\!-\!L}^4}{M_*^3}$ and
$M_{u^c_S(q_S)}\simeq  M_{\bar d^c_Hd^c_H}\simeq M_{T_1}$.
%
%
%
%
From the vectorlike heavy matter $16_f, \ov{16}_f, 10_f$ [discussed after Eq. (\ref{yuk-ops})] we assume that $n_f$ pairs of $u^c_f+\bar u^c_f$
and $l_f+\bar l_f$
happen to have masses $M_{u^c_f}$ and $M_{l_f}$, respectively, below the GUT scale $M_G$. Thus, their thresholds will also contribute to the gauge coupling running.
With all this,  imposing the unification condition for the gauge couplings at scale $M_G$, from 
renormalisation group equations we find $\al_3^{-1}(M_Z)\!=\![\al_3^{-1}(M_Z)]_{\rm MSSM}\!+\!\De_3$,
where $[\al_3^{-1}(M_Z)]_{\rm MSSM}$ corresponds to the value one would have obtained within the MSSM.
 $\De_3$ includes corrections from additional states (below the $M_G$ scale) of our model:
\begin{eqnarray}
\label{alp3}
&\De_3\!=\!\fr{3}{14\pi }\! \ln \!\!\l \!\fr{M_{D_1}\!M_{D_2}}{M_{T_2}M_{T_3}}\!\!\r^{\!\!3}\!\!\fr{M_3^8M_G^2}{M_{T_1}M_{e^c_H}^2M_8^7}\!\!\l \!\!\fr{M_G}{M_{u^c_f}}\!\!\r^{\!\!5n_f}\!\!
 \!\!\l \!\!\fr{M_G}{M_{l_f}}\!\!\r^{\!\!3n_f}\!\!\!+\!\de_3^{\rm (2)} .~~~~~
\end{eqnarray}
The $\de_3^{\rm (2)}$ stands for a two-loop correction, taken into account upon numerical calculations.
Between the thresholds' masses, the reasonable (and fairly natural)
balance, giving successful gauge coupling unification, can be found.
For instance, with the choice
\begin{eqnarray}
\label{mas-unif}
& M_{D_1}\simeq 3.6\cdot 10^{11}, ~M_{D_2}\simeq 1.05\cdot 10^{12},~M_{T_1}\!=\!10^8,
 \nonumber \\
 &M_{T_{2,3}}\!\simeq \!M_G\!=\!8\cdot 10^{16},~M_8\simeq 2.29\cdot 10^{14}, M_{3 (e^c_H)}=10^{13}, \nonumber \\
 &   M_{u^c_f}\!\simeq \!7.38\cdot \!10^{13},~ M_{l_f}\!\simeq \!1.04\cdot \!10^{15},~~ n_f\!=\!3~~
\end{eqnarray}
(all masses are given in GeV),
the gauge couplings meet at the scale $M_G\!\simeq \!8\cdot \!10^{16}$GeV, as displayed in
Fig. \ref{SO10-unif-plot}, and the unified gauge coupling is $\al_G(M_G)\simeq 0.214$ \cite{upon-anal}.
This shows how the presented scenario gives consistent gauge coupling unification.

\section{VI. Summary}
\vs{-0.3cm}
The main result of this paper is the novel mechanism suggested within the $SO(10)$ GUT
for the Higgs particle being the light pseudo-Goldstone mode.
Moreover, realizing and demonstrating the proposed mechanism, an explicit model with a fully realistic phenomenology has
been constructed.

 Within a scenario by the proposed mechanism, it is tempting to address  in  detail more phenomenological issues.
Also, it would be interesting to investigate the origin of various exploited higher-order operators through renormalizable interactions,
together with possibilities of suppressing the $V_R$ (the $45_H$'s VEV in the $I_{3R}$ direction).
For this, the proposal of Ref. \cite{Barr:1997hq} can be applied, which as shown in Ref. \cite{Babu:2010ej} (within a different construction)
can be very efficient when supported by anomalous or/and discrete symmetries.
The latter, instead of be global, can arise from discrete gauge symmetry \cite{Ibanez:1991hv} having a stringy origin.
It is tempting to use such symmetry as a flavor symmetry (gaining more motivation) and study related issues in more detail.
In this paper, I preferred to stay with the minimal setup, so I have not pursued such possibilities.
All these, supported by the mechanism discussed here,  open wide prospect towards novel $SO(10)$ GUT model building
and motivate one to study related phenomenology, such as the fermion mass pattern, neutrino oscillations, leptogenesis, and proton decay rates.
The latter still needs to be probed by ongoing nucleon decay search experiments \cite{p-decay-search}.

\section*{Appendix. Decomposition of $W_H^{(45,16)}$}

The invariant
$\l \ov{16}_H16_H\r_{\!210}\!\!\cdot \!\l 45_H^2\r_{\!210}$, which we assume to appear in the superpotential coupling $W_H^{(45,16)}$
of Eq. (\ref{WH-model}),
in explicit $SO(10)$ index contractions, is
\begin{eqnarray}
\label{210-contraction}
&\l \ov{16}_H16_H\r_{\!210}\!\!\cdot \!\l 45_H^2\r_{\!210}=\lan \ov{16}_H^T|B\Gamma_i\Gamma_j\Gamma_m\Gamma_n|16_H\ran \Phi^{[ijmn]} ) \nonumber \\
&{\rm with}~~\Phi^{[ijmn]}\!=\!45_H^{ij}45_H^{mn}\!-\!45_H^{mj}45_H^{in}\!+\!45_H^{im}45_H^{jn} .
\end{eqnarray}
Here $i,j,m,n=1-10$ and $\Ga $'s are gamma matrices in $SO(10)$'s spinorial representation and $B$ is the analog of the charge
  conjugation operator for $SO(10)$.

With $SO(10)\to SU(4)_c\tm SU(2)_L\tm SU(2)_R$ decomposition of (\ref{45-422-dec}) and
\begin{eqnarray}
\label{dec16-bar16}
&16_H=F_H(4,2,1)+F^c_H(\bar 4,1,2)~ ,\nonumber \\
&\!\! \ov{16}_H=\bar F_H(\bar 4,2,1)+\bar F^c_H(4,1,2)
\end{eqnarray}
the relevant couplings, extracted from the operator of $W_H^{(45,16)}$ in terms of $SU(4)_c\tm SU(2)_L\tm SU(2)_R$'s fragments, will be:
\begin{eqnarray}
\label{dec-W45-16-in422}
&\fr{i}{4!}\l \ov{16}_H16_H\r_{\!210}\!\!\cdot \!\l 45_H^2\r_{\!210}\to \fr{1}{2}\l \bar F_H\Si^2F_H+\bar F_H^c\Si^2F_H^c\r \nonumber \\
& -2 \bar F_H^c\Si \Si_RF_H^c+2 \bar F_H^c\Si_R^2F_H^c \nonumber \\
& -\l \fr{1}{8}{\rm tr}\Si^2+\fr{1}{2}{\rm tr}\Si_R^2\r \l \bar F_HF_H+\bar F_H^cF_H^c\r .
\end{eqnarray}
Factors in (\ref{dec-W45-16-in422}) correspond to the normalization of the VEVs:
\begin{eqnarray}
\label{VEV-normalization}
\lan \Si \ran \!=\!V_{\!B\!-\!L}\!\cdot \!{\rm Diag}\!\l 1, 1, 1, -3\r , \lan \Si_R \ran \!=\!V_{R}\!\cdot \!{\rm Diag}\!\l 1, -1\r ,~~~~~
\end{eqnarray}
where $\lan \Si \ran $ and $\lan \Si_R \ran $ are given in $SU(4)_c$ and $SU(2)_R$ group spaces respectively.
From these one can easily obtain the results presented in Eqs. (\ref{sol-S1-S})--(\ref{eq-VR}), (\ref{M-lH}), and (\ref{M-T1}).

\begin{acknowledgments}
\end{acknowledgments}
%

%


\begin{thebibliography}{26}
\bibitem{tripl-mass}
 This  issue needs additional care, because the triplet states of (\ref{M-T1}) have masses near$\sim 10^8$~GeV.
\bibitem{harmless}
 Also, there are allowed the following operators
$(S\!+\!45_H^2)S_1^310_H10_H$ and $(S\!+\!45_H^2)S_1^416_H16_H10_H$,
which do not give any relevant contribution and are safely ignored.
\bibitem{upon-anal}
Upon analysis we have taken $\al_3(M_Z)\simeq 0.1187$, and  for simplicity assumed that all SUSY states have masses$\simeq 1$TeV.
\end{thebibliography}
\end{document}